%% file: CMP_resubmission2_upload.tex
\documentclass[a4paper,aps,prx,twocolumn,nofootinbib,longbibliography,floatfix,showpacs,superscriptaddress]{revtex4-1}

\usepackage{times}
\usepackage{bbold}
\usepackage{bbm}
\usepackage[pdftex]{graphicx}
\usepackage{latexsym,amsmath}
\usepackage{boxedminipage}
\usepackage[pdftex,dvipsnames]{xcolor}
\usepackage[hidelinks,unicode=true]{hyperref}
\hypersetup{colorlinks=true,linkcolor=blue,urlcolor=blue,citecolor=blue,pdfhighlight=/N}
\usepackage{makecell}
\usepackage[normalem]{ulem}
\usepackage[english]{babel}
\usepackage{microtype}
\usepackage[normalem]{ulem}

\newcommand{\av}[1]{\langle {#1} \rangle}

\newcommand{\be}{\begin{equation}}
\newcommand{\ee}{\end{equation}}
\newcommand{\bea}{\begin{eqnarray}}
\newcommand{\eea}{\end{eqnarray}}
\newcommand{\km}{k_{min}}
\newcommand{\kmax}{k_{max}}

\begin{document}
\title{Cumulative Merging Percolation\\ and the epidemic transition of the
  Susceptible-Infected-Susceptible model in networks}

\author{Claudio Castellano}

\affiliation{Istituto dei Sistemi Complessi (ISC-CNR), Via dei Taurini
19, I-00185 Roma, Italy}

\author{Romualdo Pastor-Satorras}

\affiliation{Departament de F\'{\i}sica, Universitat Polit\`ecnica de
Catalunya, Campus Nord B4, 08034 Barcelona, Spain}

\date{\today}

\begin{abstract}
  We consider cumulative merging percolation (CMP), a long-range percolation
  process describing the iterative merging of clusters in networks,
  depending on their mass and mutual distance.  For a specific class of CMP
  processes, which represents a generalization of degree-ordered
  percolation, we derive a scaling solution on uncorrelated complex
  networks, unveiling the existence of diverse mechanisms leading to the
  formation of a percolating cluster.  The scaling solution accurately
  reproduces universal properties of the transition.  This finding is used
  to infer the critical properties of the Susceptible-Infected-Susceptible
  (SIS) model for epidemics in infinite and finite power-law distributed
  networks. Here discrepancies between analytical approaches and
    numerical results regarding the finite size scaling of the epidemic
    threshold are a crucial open issue in the literature.  We find that the
    scaling exponent assumes a
    nontrivial value during a long preasymptotic regime. We calculate this
    value, finding good agreement with numerical evidence.  We also show
    that the crossover to the true asymptotic regime occurs for sizes much
    beyond currently feasible simulations. Our findings allow us to
    rationalize and reconcile all previously published results (both
  analytical and numerical), thus ending a long-standing debate.
\end{abstract}

\maketitle

\section{Introduction}
\label{sec:introduction}

Percolation and epidemic spreading are among the most interesting processes
unfolding on complex network substrates and their investigation has
attracted a huge interest in the past 20
years~\cite{Newman10,dorogovtsev07:_critic_phenom,Pastor-Satorras:2014aa,porter2016dynamical,Kiss2017}.
One of the most successful achievements of this endeavor is the realization
that the properties of one of the fundamental models for epidemics without a
steady state, the susceptible-infected-recovered (SIR)
dynamics~\cite{epidemics}, can be mapped onto bond
percolation~\cite{Grassberger1983,Newman2002}~\footnote{Bond percolation is
  exactly mapped to the final state of the Independent Cascade Model, a
SIR-like process having fixed recovery time.  For SIR the mapping is on a
semi-directed epidemic percolation network, as explained in
Ref.~\cite{Kenah2007}.}. This connection has permitted the application to
the SIR model of the powerful tools devised for percolation, leading to a
full understanding of this epidemic process~\cite{Newman2002, Kenah2007,
PhysRevLett.97.088701, Karrer2010, Gleeson2018}.  For the other fundamental
class of epidemic dynamics, allowing for a steady, endemic state, whose
simplest representative is the susceptible-infected-susceptible (SIS)
model~\cite{epidemics}, no direct mapping to a percolative framework is
available and theoretical progress has been slower. In the SIS model,
susceptible individuals acquire the disease at rate $\beta$ through any edge
connected to an infected individual, while infected individuals
spontaneously heal with rate $\mu$. The epidemic threshold $\lambda_c$,
defines the value of the ratio $\lambda = \beta / \mu$ separating a healthy
(absorbing) phase from an endemic one with everlasting activity.  Initial
work showed that degree heterogeneity leads to disruptive effects on
scale-free networks~\cite{Barabasi:1999}, namely, a vanishing threshold in
networks with power-law degree distribution $P(k) \sim k^{-\gamma}$ and
$\gamma\leq 3$~\cite{pv01a,Pastor01b}. Later efforts have shifted toward
less heterogeneous networks, those with
$\gamma>3$~\cite{Pastor-Satorras:2014aa}.

The Quenched Mean-Field (QMF)
theory~\cite{Wang03,PVM_ToN_VirusSpread,Gomez10} predicts a vanishing
threshold $\lambda_c \to 0$ in the infinite network-size limit for any value
of $\gamma$~\cite{Castellano2010}, due to the existence of hubs able to
sustain the epidemic for long times only by interacting with their direct
neighbors~\cite{Castellano2012}.  It was later pointed out that, at the QMF
level, the localization of activity around these hubs implies the existence,
for small values of $\lambda$, of long-lived, but not stationary,
states~\cite{Goltsev2012,Lee2013}.  An important progress in this debate was
provided in Ref.~\cite{Boguna2013}, where it was shown that a genuine non
mean-field effect, mutual reinfection among distant hubs, is the key
mechanism triggering the appearance of an endemic stationary state for any
$\lambda$ in networks with $\gamma>5/2$.  Numerical evidence corroborated
this picture, showing that the position of the effective threshold tends to
zero with network size for any $\gamma$. However, the decay observed
  was slower than the one predicted by QMF
theory~\cite{Boguna2013,Mata2015,Arruda2018} and moreover in contradiction
with recent mathematical results derived by Huang and 
Durrett~\cite{Huang2018} and
Mountford et al.~\cite{Mountford2013}.   An additional puzzling question in
this area is the striking disagreement between the exact mathematical
prediction for the singular behavior of the prevalence ($\rho \sim
\lambda^{2\gamma-3}$, apart from logarithmic
corrections)~\cite{Mountford2013} and numerical simulations exhibiting a
much faster growth. This lack of a precise agreement between analytics
  and numerics represents one standing issue in our understanding of
  epidemic processes on complex topologies.

A precise mathematical formulation of the mutual reinfection process was
recently proposed by M\'enard and Singh~\cite{Menard2015}. They introduced
the cumulative merging percolation (CMP) process, a long-range site
percolation process~\cite{grimmett1999} aimed at describing the geometry of
the sets where SIS epidemics survives for a long time on a network.  The
presence of a CMP giant component corresponds to the existence of an endemic
SIS stationary state, so that the calculation of the CMP threshold allows
one to locate also the position of the SIS epidemic transition\footnote{In
Ref.~\cite{Menard2015} it is demonstrated that the CMP threshold is a
lower-bound for the epidemic threshold. Based on the physical picture, we
expect the two quantities to coincide.}.

In this paper we contribute to the current state-of-the-art in this area in
two ways.  First, we consider a generalized version of the CMP process
proposed in Ref.~\cite{Menard2015} and we present a scaling theory for its
nontrivial behavior. This theory -- which provides a clear understanding of
competing physical mechanisms, critical properties, crossover scales and
finite size effects -- is general and can be related with other processes.
Second, we apply the results of the first part to SIS dynamics, obtaining in
this way for the first time a full understanding of the critical properties
of the model.  In particular, our theory predicts that the asymptotic
behavior in the limit of very large networks (derived exactly
in~\cite{Huang2018,Mountford2013} and
reassuringly recovered by our approach) can be observed only for huge system
sizes, out of reach for present computer resources.  We show instead that,
for network sizes that can be currently simulated, a preasymptotic regime
holds, whose nontrivial properties are determined, providing a prediction
for the finite size scaling of the SIS epidemic threshold in agreement with
(previously unexplained) numerical results.  Our work reconciles in a
comprehensive way the different theories proposed to interpret the behavior
of the SIS model, placing them in the proper context regarding the network
size considered, and thus ends a long debate between the physics and
mathematics communities.

The paper is organized as follows: In Sec.~\ref{sec:gener-cumul-perc} we define
the cumulative merging percolation process which will be the subject of our
study. Sec.~\ref{sec:scal-theory-gener} presents a scaling solution of this
model, whose behavior in finite networks is discussed in
Sec.~\ref{sec:finite-size-effect}. A numerical check of the scaling solution is
provided in Sec.~\ref{sec:numerical-test}.  In Sec.~\ref{sec:appl-sis-epid} we
apply the results obtained to the SIS epidemic model, backing up our conclusions
by comparison with existing numerical simulations. Finally, in
Sec.~\ref{sec:discussion} we summarize our main results and discuss the
interesting perspectives they open.  Several appendices provide some detailed
analytical calculations and additional information.

\section{Cumulative Merging Percolation Process}
\label{sec:gener-cumul-perc}

\begin{figure}[t]
  \includegraphics*[width=\columnwidth]{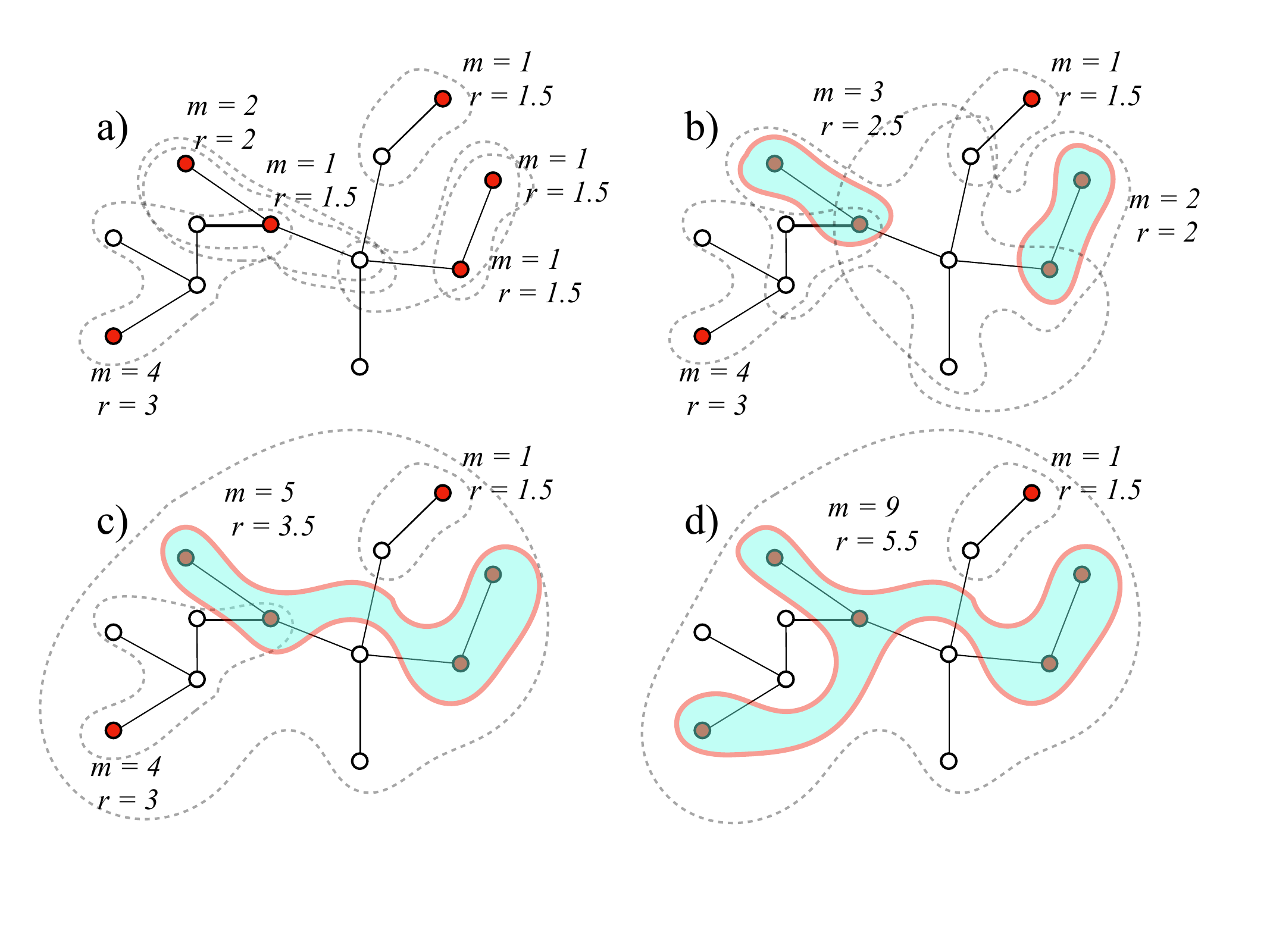}
  \caption{Schematic illustration of the CMP process for $r(m)=1+m/2$.  Filled
    nodes are active, empty nodes are inactive. Areas bordered by dashed lines
    are interaction domains of active nodes or clusters.  Clusters are indicated
    by solid lines surrounding a filled area.  Notice in panels c) and d) the
    isolated active node on the upper right corner, which is within the
    interaction range of the large cluster but cannot be merged as it has
    $r<2$.}
\label{figS1}
\end{figure}

We consider a generalization of the cumulative merging process proposed in
Ref.~\cite{Menard2015}, defined along the following lines.  In
a given network,
composed by $N$ nodes, each node $i$ is \textit{active} with probability
$p_i$. Inactive nodes do not play any role apart from determining the
topological distances between pairs of active nodes (see below).  Each active
node $i$ defines a cluster of size 1, associated with an initial mass
$m_i^{(0)}$.  Starting with these initial clusters, an iterative process takes
place whose elementary step is the merging of a pair of clusters into a single
one.  Two clusters, $\alpha$ and $\beta$, are merged in a single cluster if
there are at least a node $i_\alpha$ in $\alpha$ and 
a node $j_\beta$ in $\beta$, such that
\begin{equation}
  d_{i_\alpha,j_\beta} \le
  \min\{r(m_\alpha), r(m_\beta)\},
  \label{CMPcondition}
\end{equation}
where $d_{i,j}$ is the topological distance between nodes $i$ and $j$, and
$r(m) \geq 1$
is an interaction range associated to a cluster of mass $m$.  The
mass of the merged cluster is the sum of the masses of the original clusters,
$m_{\alpha+\beta}=m_\alpha+m_\beta$.  The iteration of this procedure converges
to a limiting partition of the network that does not depend on the order in
which the merging is performed\footnote{Although this is mathematically proved
  only in a specific case in Ref.~\cite{Menard2015}, we found numerically the
  same independence in the cases considered below.}.  Notice that if $p_i=p$ and
$r(m)=1$ CMP coincides with random site percolation~\cite{Newman10}.  It is
important to remark that Eq.~(\ref{CMPcondition}) implies that two clusters
merge only if each one of them is within the interaction range of the other: An
asymmetric situation, with a massive cluster interacting with a far and small
cluster, does not lead to merging. In Fig.~\ref{figS1} we present a graphical
illustration of the mechanism of the CMP process.
We stress again that a cluster is defined as a set of (only) 
active nodes resulting from the iteration of merging events. 
Nodes in the same cluster must belong to the same connected component
of the underlying network, but they do {\em not} need to form 
a connected component by themselves, 
as it is clear from panels c) and d) of Fig.~\ref{figS1}.

The connection between CMP and the mutual reinfection of distant hubs
in the SIS epidemics is operated by taking as active nodes the hubs
able to independently sustain the epidemic~\cite{Menard2015}, see
Appendix~\ref{sec:conn-betw-cmp} for a detailed description.

\section{Scaling theory for Cumulative Merging Percolation}
\label{sec:scal-theory-gener}

Let us focus now on a specific yet broad class of CMP processes, where nodes are
active if their degree is larger than a threshold value $k_a$,
$p_i\equiv p(k_i)=\Theta(k_i-k_a)$.  In an uncorrelated network with
degree distribution $P(k)=(\gamma-1)\km^{\gamma-1} k^{-\gamma}$ in the
continuous approximation, where $\km$ is the minimum degree, the fraction of
active nodes is
\begin{equation}
  \label{eq:2}
  \frac{N_a}{N} = \int_{k_a}^{\infty} dk P(k) =
  \left(\frac{k_a}{\km}\right)^{1-\gamma}.
\end{equation}
We are interested in understanding the possible existence of a CMP giant component
as a function of $k_a$, in particular in the limit $k_a \to \infty$,
when only a small fraction of nodes is active.

\subsection{The case $r(m)=1$: Degree-Ordered Percolation}

Let us consider first the case $r(m) = 1$, i.e., only nearest
neighbors can form clusters.
In this case the CMP process defined above coincides with the
degree-ordered percolation (DOP) process proposed in Ref.~\cite{Lee2013}
(coinciding with the limit $\alpha \to -\infty$ in Ref.~\cite{Gallos2005}).
For a node of degree $k$, the probability that a given neighbor is active is
\begin{equation}
  P_a(k) = \int_{k_a}^{\infty} dk' P(k'|k),
\end{equation}
where $P(k'|k)$ is the conditional probability that a neighbor of a node $k$ has
degree $k'$~\cite{alexei}. For uncorrelated networks
$P(k'|k) = \frac{k' P(k')}{\av{k}}$~\cite{alexei}, thus we have
$P_a = \left(\frac{k_a}{\km}\right)^{2-\gamma}$, independent of $k$.  The mean
number of active neighbors of a node of degree $k$ is $k P_a$; therefore the
inverse of $P_a$,
\begin{equation}
  k_c = \left(\frac{\km}{k_a}\right)^{2-\gamma},
\end{equation}
defines a degree scale separating nodes likely to have many active neighbors
$k/k_c \gg 1$ from those likely to be {\em isolated}, i.e., not in direct
contact with any active node.  The average number of active neighbors for each
active node is
\begin{equation}
  \frac{N}{N_a}\int_{k_a}^\infty dk P(k) k P_a =
  \frac{\gamma-1}{\gamma-2} k_a P_a \sim k_a^{3-\gamma}.
  \label{aveka}
\end{equation}
For $\gamma<3$ this quantity diverges as $k_a$ grows: Each active node
has a very large number of active neighbors, so that all of them belong
to a connected giant component for any $k_a$~\cite{Lee2013,Gallos2005},
and the relative  size $S$ of the giant component is simply given by
the fraction of active nodes
\be
S_{DOP} = \frac{N_a}{N}  = \left(\frac{k_a}{\km}\right)^{1-\gamma}.
\label{SDOP}
\ee

For $\gamma>3$, instead, the average number of active neighbors of an active
node decreases with $k_a$ and tends to zero in the limit $k_a \to \infty$.  This
indicates that a degree-ordered percolation giant component (DOPGC) can exist
only up to a finite threshold value, in agreement with
Refs.~\cite{Lee2013,Gallos2005}.
It is useful to discuss the behavior of the order parameter $S_{DOP}$ as a
function of $k_a$ in this case.  For $k_a=\km$, $k_c=1$. Hence, even for
$\gamma>3$, there is an interval of $k_a$ values such that $k_a/k_c>1$.  This
regime occurs up to a value $k_a=k_0^*$ determined by the condition
$k_c(k_0^*) = k_0^*$, yielding
\begin{equation}
  k_0^* = \km^{(\gamma-2)/(\gamma-3)}.
\end{equation}
Notice that for $\gamma=3.2$ and $\km=3$, $k_0^*=729$, a quite large value,
while it decays quickly for increasing $\gamma$: For $\gamma=3.5$ it is already
$k_0^*=27$.  In this regime the situation is similar to the case $\gamma<3$,
with practically all active nodes belonging to the DOPGC and
$S_{DOP} \approx N_a/N \sim k_a^{1-\gamma}$.  However, one must notice that,
even if $k_a/k_c>1$, this ratio is not very large, as its maximum value is
$\km$, corresponding to $k_a = \km$.
Therefore, one never observes the scaling
predicted by Eq.~(\ref{SDOP}); as soon as $k_a$ is increased one immediately
starts to see the transition to a different regime, where $k_a/k_c < 1$.  In
this second regime a giant component still exists, but some active nodes are
isolated (not directly connected to other active nodes) and others are
non-isolated but form small clusters.  The set of all active nodes is therefore
composed by three classes:
\begin{enumerate}
\item
  Non-isolated nodes belonging to the DOPGC;

\item
  Non-isolated nodes belonging to small clusters;

\item
  Isolated nodes, which necessarily do not belong to the DOPGC.
\end{enumerate}

\begin{figure}[t]
  \includegraphics[width=\columnwidth]{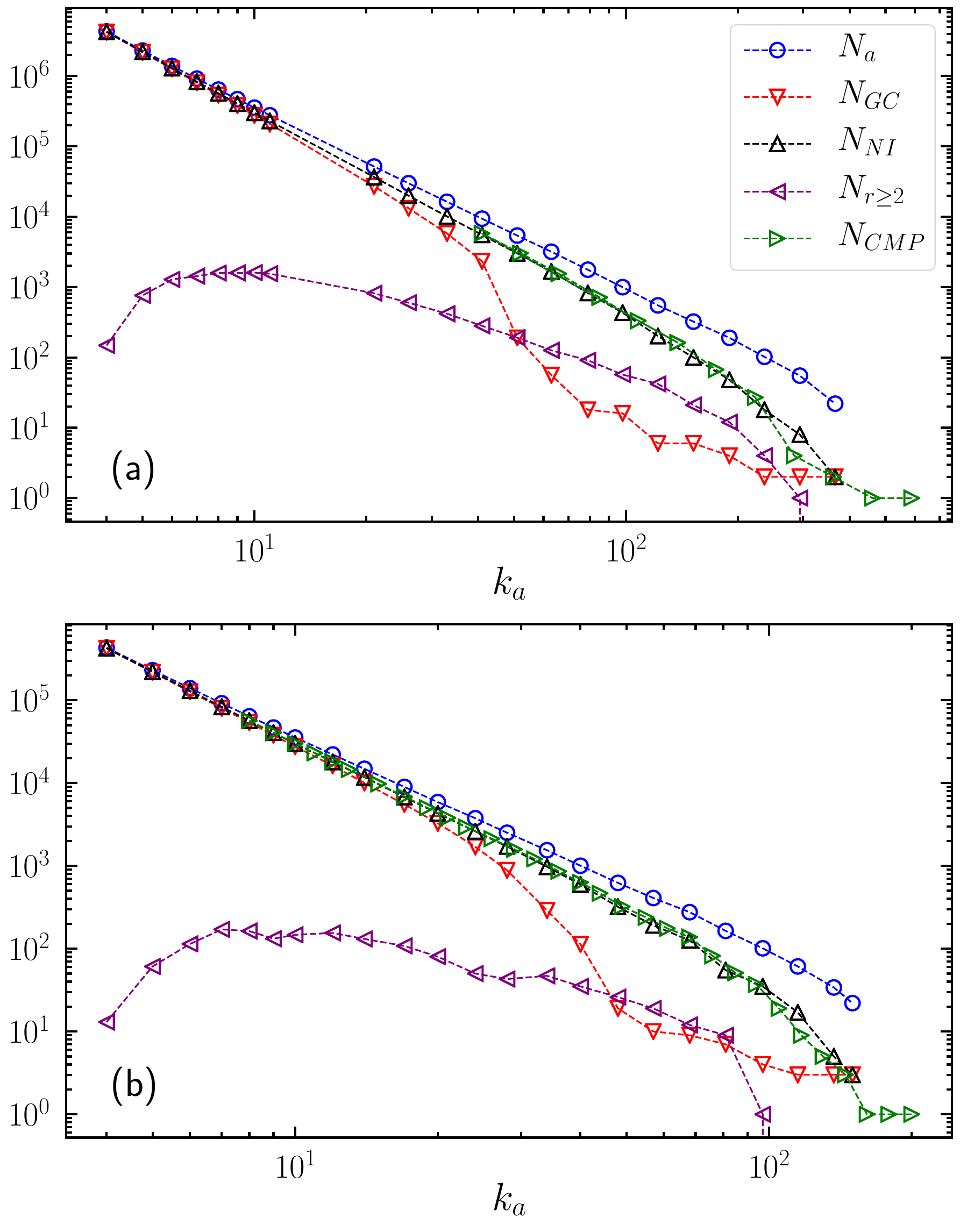}
  \caption{Dependence on $k_a$ of
    the number $N_a$ of active nodes, the number
    $N_{GC}$ of nodes in the DOP giant component, the number $N_{NI}$ of
    nonisolated nodes, the number $N_{r\geq2}$ of isolated nodes with range
    $r \geq 2$, and the number $N_{CMP}$ of nodes in the CMP giant component.
    Results are for power-law networks with $\gamma=3.5$, $\km=3$
    and size $N=10^7$ (a)
    and $N=10^6$ (b), generated using the uncorrelated configuration model
    (UCM)~\cite{Catanzaro2005}.}
\label{figDOP}
\end{figure}

As $k_a$ increases, a growing fraction of active nodes passes from the first
category to the other two, and the order parameter $S_{DOP}=N_{GC}/N$ decreases
faster than the fraction of active nodes $N_a/N$ (see Fig.~\ref{figDOP}).  At
the threshold the fraction of non-isolated nodes belonging to the DOPGC vanishes.

The calculation of the behavior of $S_{DOP}$ in this regime and of the
transition point is a nontrivial task.  It is important to observe that
$N_{NI}$, the number of nonisolated nodes, which upper bounds the number
$N_{GC}$ of nodes belonging to the giant component, keeps decaying with the same
exponent even well above the DOP transition (see Fig.~\ref{figDOP}).

\subsection{The case of growing $r(m)$ for $\gamma<3$}

Let us turn now to the more generic case where $r(m)$ grows as a function of
$m$. The range of interaction grows with its mass, so that, if $r(m) \ge 2$,
clusters of nodes can merge even if not in direct contact.  In this case, it is
clear that, for a given value of $k_a$, the giant component of the DOP process
is a subset of the giant component of the full CMP process (CMPGC).  Thus, for
$\gamma<3$, the CMPGC is again given by the whole set of active nodes, and has
therefore a relative size 
\be S = \frac{N_a}{N} =
\left(\frac{k_a}{\km}\right)^{1-\gamma}.
\label{S}
\ee

\subsection{The case of growing $r(m)$ for $\gamma>3$}

In this case, for large $k_a$ the DOPGC vanishes asymptotically and non-isolated
active nodes form DOP clusters of small size.  Still an extensive CMPGC could be
induced by long range merging of clusters or nodes which cannot be joined in a
DOP process, as they are separated by distances larger than $1$.  Whether these
long range mergings take place or not depends of course on the particular choice
of the mass $m$ and of the form of the interaction range.  Inspired by
Ref.~\cite{Menard2015}, here we focus on the case of initial masses equal to
node degrees $m_i^{(0)} = k_i$ (so that the total mass of a cluster is the sum
of the degrees of the active nodes forming it) and of an interaction range of
the form $r(m) = m/k_a$.  This is a particular case of a generic CMP process
with $r(m)=f(m/k_a)$ where $f(z) = z^{\alpha}$, with $\alpha>0$, so that active
nodes with the smallest degree have range exactly equal to $1$.  We defer to a
future work a comprehensive analysis of this model for $\alpha \ne 1$.

In the present setting we identify two competing mechanisms leading to the
formation of a CMP giant component.  The first is an extension of DOP
percolation, based on the merging of DOP clusters separated by distances larger
than 1.  The second involves the buildup of CMP clusters formed by isolated
nodes interacting at large distance.  We now discuss the two mechanisms in
detail.

\subsubsection{First mechanism: Extended DOP mechanism}

For very small $k_a$ close to $\km$, CMP is clearly equivalent to the first
regime for DOP with $S \approx N_a/N$.  Upon increasing $k_a$, above the
crossover scale $k_0^*$, DOP enters the second regime with an increasing
presence of isolated nodes and nodes belonging to small DOP clusters.  CMP and
DOP behaviors start to diverge at this point because some nodes, even if they
are not directly connected to the DOPGC, they are at distance 2 from it and thus
can join the CMPGC if their interaction range is at least 2.  In particular this
occurs for all small DOP clusters: As their aggregate degree is
$k_{agg} \ge 2 k_a$ they necessarily have a range of interaction $r \ge 2$. For
this reason, in this regime all $N_{NI}$ non-isolated nodes belong to the CMPGC.
This is clearly verified in Fig.~\ref{figDOP}.  Notice that $N_{NI}/N$ is finite
even well beyond the DOP threshold.  In this limit, the formation of the CMPGC is
still triggered by the largest DOP cluster (that does not percolate). For any
value of $\gamma$ there are always nodes in the network with $k> k_c \gg k_a$.
They form local clusters with large interaction range that progressively
incorporate other small clusters giving rise to a CMPGC, even if no DOPGC is present.
To calculate $N_{NI}$, we consider the probability that an active node of degree
$k$ has at least one neighboring active node
\begin{equation}
  P_{NI}(k) = 1-(1-P_a)^k \approx 1-e^{-k/k_c}.
  \label{P1}
\end{equation}
The total fraction $N_{NI}/N$ of non-isolated active nodes in a
power-law distributed network is then
\begin{eqnarray}
\frac{N_{NI}}{N}&=& \int_{k_a}^{\infty} dk' P(k') P_{NI}(k') \label{eqNlc}
  \\ \nonumber
  &= & (\gamma-1)
       \km^{\gamma-1} \left[\frac{k_a^{1-\gamma}}{\gamma-1}-k_c^{1-\gamma}
       \Gamma\left(1-\gamma,\frac{k_a}{k_c}\right) \right],
\end{eqnarray}
where $\Gamma(a, z)$ is the incomplete Gamma function~\cite{abramovitz}.

In this second regime, not only small DOP clusters, but also isolated active
nodes can join the CMPGC, provided they have degree $k \ge 2 k_a$ so that their
range is $r \ge 2$.  We denote their number as $N_{r\ge2}$.  The total fraction
of isolated nodes with range $r \ge 2$ is
\begin{eqnarray}
  \frac{N_{r\ge2}}{N} &=& \int_{2 k_a}^{\infty} dk' P(k') [1-P_{NI}(k')] \\
                      &=& (\gamma-1) \km^{\gamma-1} k_c^{1-\gamma}
                          \Gamma\left(1-\gamma, \frac{2k_a}{k_c} \right).
\end{eqnarray}
Overall the CMP order parameter in this regime is therefore
\begin{equation}
  S_1 \approx \frac{N_{NI}}{N}+\frac{N_{r\ge2}}{N}.
  \label{S11}
\end{equation}

For $k_a \to \km$ one has $k_a > k_c$ and the first contribution in
Eq.~(\ref{S11}) is larger than the second, for any $\gamma$.
For large $k_a$ instead, one can expand the $\Gamma$ functions for small
$k_a/k_c$, finding
\begin{eqnarray}
\frac{N_{NI}}{N} = \left(\frac{k_a}{\km}\right)^{1-\gamma}
\left[\frac{\gamma-1}{\gamma-2} \frac{k_a}{k_c} \right] \sim
k_a^{2(2-\gamma)}
\end{eqnarray}
and
\begin{eqnarray}
  \frac{N_{r \ge 2}}{N} =  \left(\frac{2 k_a}{\km}\right)^{1-\gamma}
  \left[ 1  -\frac{\gamma-1}{\gamma-2}
\frac{2 k_a}{k_c} \right] \sim k_a^{1-\gamma}.
\end{eqnarray}
The exponent of $N_{NI}$ is, in absolute value, larger than the one of
$N_{r \ge 2}$, hence the first contribution dominates up to a crossover scale
\begin{equation}
  k_1^* =
  \left[\frac{(\gamma-2)}{(\gamma-1)}\frac{2^{(1-\gamma)}}
    {(1+2^{2-\gamma})} \km^{(2-\gamma)} \right]^{1/(3-\gamma)}.
  \label{k1}
\end{equation}

The conclusion of this line of reasoning is that for $k_a \ll k_1^*$
the size of the CMPGC decays as
\begin{equation}
  S_1 \approx \frac{N_{NI}}{N} \sim k_a^{2(2-\gamma)}
  \label{S1}
\end{equation}
followed by a crossover to
$S_1 \approx \frac{N_{r \ge 2}}{N} \sim k_a^{1-\gamma}$.  The crossover scale
$k_1^*$ decreases rapidly with $\gamma$ but, since the maximum degree in a
network grows as $N^{1/(\gamma-1)}$, the minimum network size necessary to have
a sufficiently large maximum degree $\kmax=k_1^*$ is always larger than
$N \approx 4.2 \times 10^5$ (the minimum occurring for $\gamma \approx 5$ for
$\km=3$).  Hence it should be possible to observe
the crossover on large networks (although $\kmax$ grows very slowly with $N$,
hence one needs networks of size much larger than $10^5$ nodes to have a still
limited range of $k_a$ values).  As a matter of fact, we do not observe such a
crossover.

\begin{figure}
  \includegraphics[width=\columnwidth]{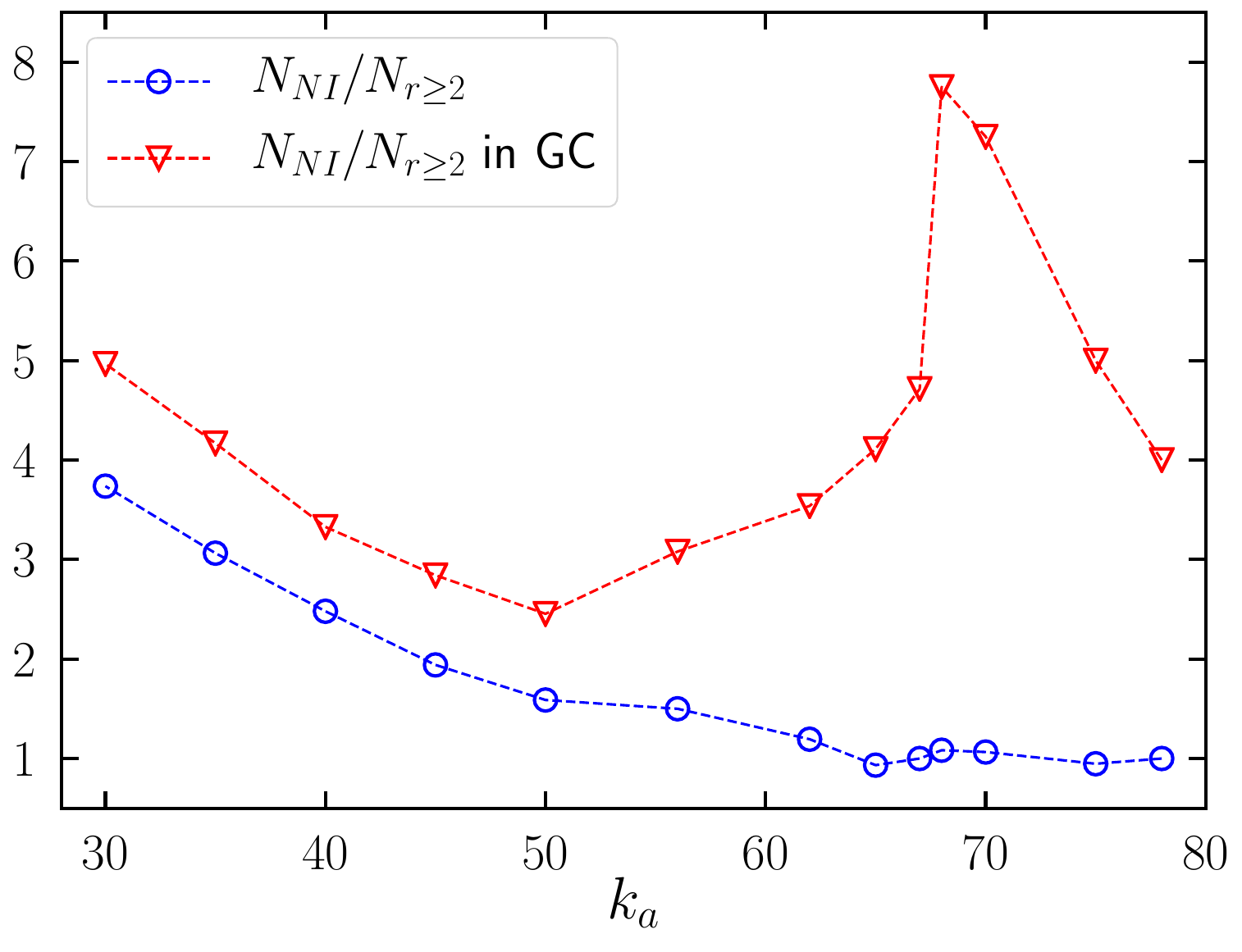}
  \caption{Ratio $N_{NI}/N_{r\ge2}$ between the two terms in Eq.~(\ref{S11})
    evaluated on UCM networks with $\gamma=4$, $\km=3$ and size $N=10^7$, computed over
    all nodes, and restricted to nodes belonging to the CMPGC.}
  \label{figRatio}
\end{figure}

This happens because, as $k_a$ grows, the extended DOP mechanism
becomes less and less effective. DOP clusters become smaller and
smaller and the distances among them (and between isolated active nodes
and them) increase: it is no more sufficient to have $r=2$ to join the
CMP giant component. For even larger $k_a$ it is not even sufficient
to have $r=3$ or $r=4$ and so on.
This effect suppresses both terms in Eq.~(\ref{S11}), but the second
term is most affected, as can be seen in Fig.~\ref{figRatio}, where
we compare the ratio of first and the second term in
Eq.~(\ref{S11}) (which becomes 1 at the crossover scale $k_1^*$) and the same
ratio restricted to nodes belonging to the CMPGC. We observe that the latter is
always larger than the former and does not seem to go to 1 for large $k_a$.
This implies that in practice $S_1$ behaves as
predicted by Eq.~(\ref{S1}) even for values of $k_a$ larger than the
crossover scale $k_1^*$ estimated in Eq.~(\ref{k1}).

A more important consequence of the asymptotic ineffectiveness of
the extended DOP mechanism is that it cannot work for arbitrarily large $k_a$.
A different mechanism governs the formation of the CMPGC in the
limit $k_a \to \infty$.

\subsubsection{Second mechanism: Merging of distant isolated nodes}

Nodes with degree $k_a \le k \ll k_c$ have on average a
very small number of active nearest neighbors, as $k P_a = k/k_c \ll 1$.
Hence they are typically isolated.
However, if $k$ is large enough, they may still have a large interaction
range and may merge with other distant nodes. To analyze this process in
detail, let us denote as $d(k)$ the mean distance between a node of degree
$k$ and the closest node of degree at least $k$.
In the limit of large network size, this distance is
(see Appendix~\ref{sec:aver-dist-betw} for an analytical derivation)
\begin{equation}
  d(k) \approx 1+ \frac{\gamma-3}{\ln(\kappa)}\ln\left(\frac{k}{\km}\right),
\end{equation}
where $\kappa = \av{k^2}/\av{k}-1$ is the network branching factor.  Since the
interaction range of a node grows linearly with its degree $k$, it grows faster
than the distance to its closest peer.  Hence there exists a degree $k_x$ such
that
\begin{equation}
  r(k_x)=d(k_x).
  \label{rkx}
\end{equation}
and for any $k > k_x$, $r(k) > d(k)$.  As a consequence nodes with $k > k_x$
have an interaction range larger (on average) than their mutual topological
distance.  They can thus merge in pairs with an even larger interaction range
and the process repeats itself leading to the formation of a CMPGC, comprising
all nodes with degree larger than $k_x$.  If we write $k_x$ in the form
$k_x=\omega k_a$ the condition~(\ref{rkx}) implies $\omega = d(\omega k_a)$,
which, inserting the explicit expression of $d(k)$, becomes
\begin{equation}
  \omega \approx 1+
  \frac{\gamma-3}{\ln(\kappa)}
  \ln \left(\frac{\omega k_a}{\km} \right)
\label{eqomega}
\end{equation}
Neglecting constants and terms of order $\ln[\ln(k_a)]$, the size of the giant
component according to this mechanism scales then as
\begin{eqnarray}
S_2 &\approx &\left( \frac{k_x}{\km} \right)^{1-\gamma} =
\omega^{1-\gamma} \left( \frac{k_a}{\km} \right)^{1-\gamma} \\
& = & \left[\frac{\gamma-3}{\ln(\kappa)}
\ln \left(\frac{k_a}{\km} \right)
\right]^{(1-\gamma)}
\left(\frac{k_a}{\km} \right)^{1-\gamma}
\label{S2}
\end{eqnarray}
showing thus a power-law decay times a logarithmic correction.

In absolute value, the leading exponent in $S_2$ is smaller than the exponent
in $S_1$.  Therefore we expect this second mechanism to dominate asymptotically,
but after a crossover preceded by a scaling regime where the size
of the CMPGC is given by Eq.~(\ref{S1}).
The position $k_2^*$ of the crossover
is estimated by numerically solving the equation $S_1(k_2^*) = S_2(k_2^*)$.
\begin{figure}
  \includegraphics[width=\columnwidth]{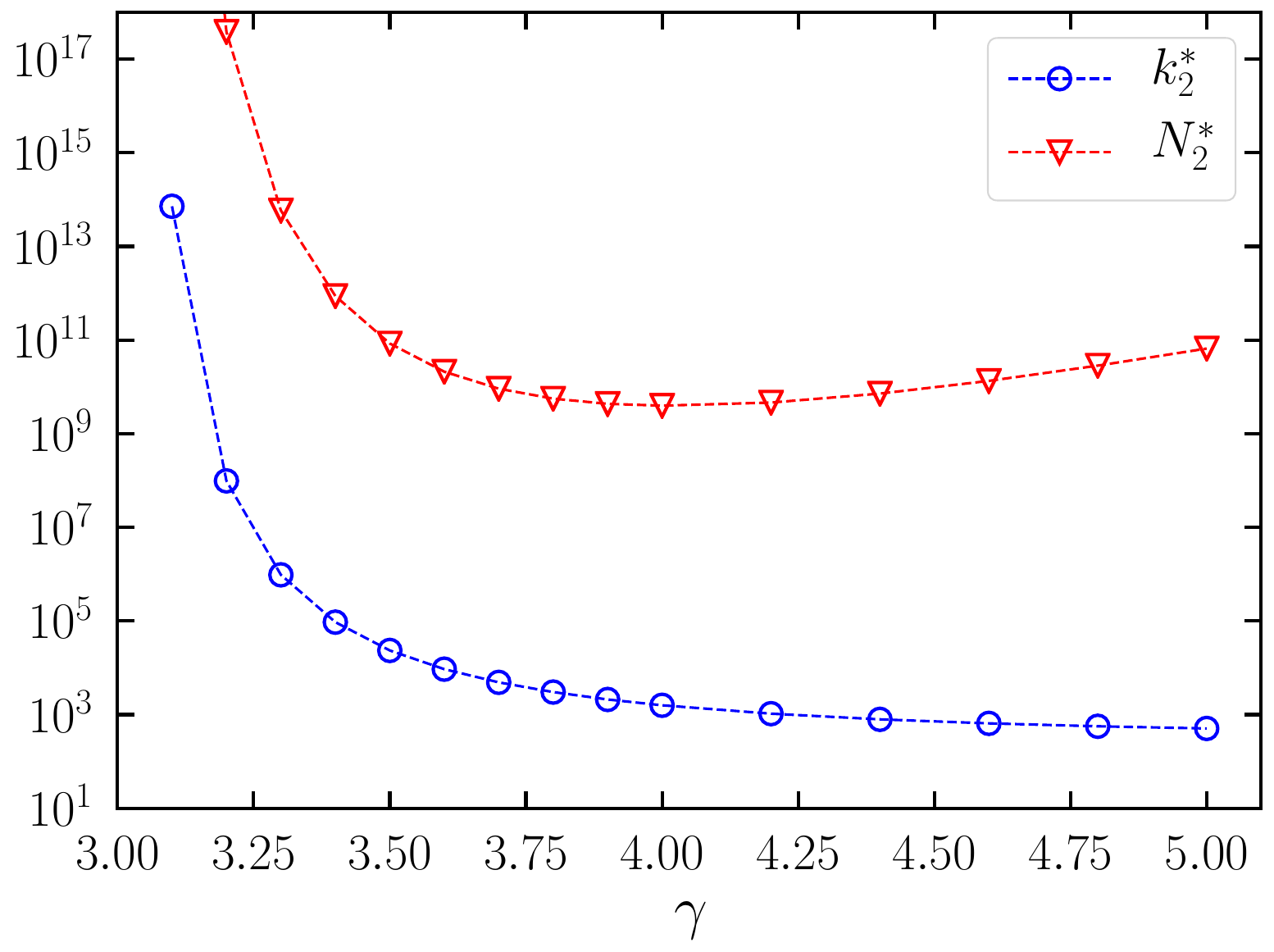}
  \caption{Values of the crossover degree $k_2^*$ and the minimal network size
    $N_2^*=k_2^{*(\gamma-1)}$ as a function of $\gamma$.  In order to observe
    the crossover, networks of size $N \gg N_2^*$ should be considered.}
  \label{figCrossover}
\end{figure}
Fig.~\ref{figCrossover} shows how this quantity decreases with the exponent
$\gamma$. However, in order to observe such a crossover one must consider
networks much larger than $N_2^*=k_2^{*(\gamma-1)}$.
These values are huge for any $\gamma$ (much larger than
$10^9$ nodes in the best case),  leading to the conclusion that only
the first regime can be observed in currently feasible simulations.

The present analysis can be extended also to the case of
networks with a stretched exponential degree distribution,
predicting an asymptotic stretched exponential dependence of
$S$ on $k_a$.
See Appendix~\ref{stretched} for details.

\section{Finite-size effects}
\label{sec:finite-size-effect}

So far we have considered infinitely large networks, thus assuming that
all degree classes, up to infinity, exist.
When the network size is finite, only degrees up to the maximum value
$\kmax(N)$, growing as $N^{1/(\gamma-1)}$, are present~\cite{mariancutofss}.
The CMP behavior for the infinite network (i.e. there is a
CMPGC for any $k_a$) holds as long as $\kmax$ is larger than the degree
scale involved in the formation of the CMPGC.

For $\gamma<3$, it is sufficient to have active nodes for observing an extensive
CMPGC. Hence the only finite size effect trivially appears for $k_a>\kmax(N)$:
In such a case there are no more active nodes in the system and $S \approx 0$.
The finite size effective threshold is $k_a^c=\kmax(N)$.

On the contrary, for $\gamma>3$ finite size effects are less trivial.  The
presence of active nodes is not sufficient to give rise to a CMPGC.  One needs
the presence of nodes with $k>k_c$ (first mechanism) or $k>k_x$ (second mechanism).
Notice that since $k_c$ grows as a power of $k_a$ with exponent larger than 1,
while $k_x$ grows logarithmically, asymptotically $k_c \gg k_x$.
Different scalings of the finite-size effective threshold are
possible, depending on whether the maximum degree $\kmax(N)$ is larger or smaller
than the crossover degree $k_2^*$.

If $\kmax(N) > k_2^*$, finite size effects appear during the
regime where the formation of the CMPGC is governed by the second
mechanism.
In this case the asymptotic behavior $S \approx S_2$ ends (i.e., $S \approx 0$)
when the relevant degree scale $k_x$ (growing with $k_a$) becomes
larger than $\kmax(N)$.
In such a case there are active nodes in the system, but neither of
the two mechanisms for the formation of the giant component is at work.
The effective threshold in this case is given
by the condition $k_x = \kmax(N)$, implying asymptotically
\begin{equation}
  k_a^c \sim \frac{\ln (\kappa)}{\gamma-3} \frac{\kmax(N)}{\ln(\kmax(N))}.
\label{k_ac2}
\end{equation}

If instead $\kmax(N) < k_2^*$, finite size effects start to appear
already during the preasymptotic regime where the first mechanism rules.
As soon as $k_c>\kmax(N)$, the behavior $S \approx S_1$ ends.
The effective threshold is thus given by the condition $k_c=\kmax(N)$,
implying:
\begin{equation}
  k_a^c = \km \kmax^{1/(\gamma-2)}.
  \label{k_ac}
\end{equation}
Notice that after this effective threshold the order parameter does not go
to $S \approx 0$, as
there is still an interval of $k_a$ values such that $k_x < \kmax(N) < k_c$.
In this regime the first mechanism is no more operative; still the second
is at work, but since $N < N_2^*$, it cannot lead to a macroscopic giant component.

\section{Numerical test}\label{sec:numerical-test}

\begin{figure}
  \includegraphics[width=\columnwidth]{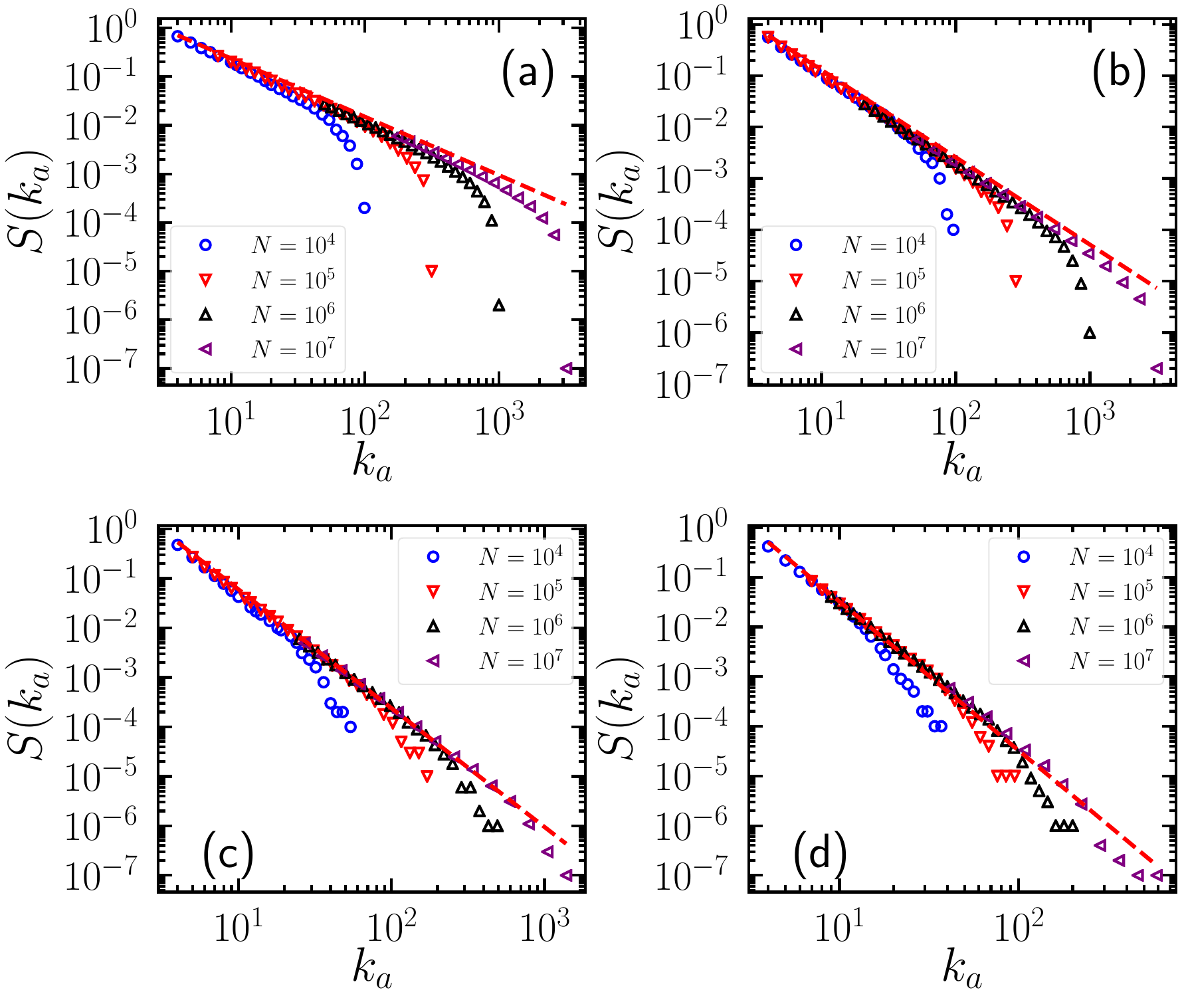}
  \caption{Fraction $S$ of nodes in the largest CMP component as a function of
    $k_a$ for various $\gamma$ values: $\gamma=2.2$ (a), $\gamma=2.7$ (b),
    $\gamma=3.2$ (c), $\gamma=3.5$ (d). In all cases $\km=3$.
    Symbols represent numerical results for
    various network sizes.  Dashed lines are theoretical predictions from
    Eq.~(\ref{S}) [panels (a) and (b)] and Eq.~(\ref{S1}) [panels (c) and (d)].}
  \label{figNNI}
\end{figure}

We test the correctness of the scaling analysis performed in the previous
Sections by means of numerical simulations of the CMP process with $r(m)=m/k_a$.
In Figs.~\ref{figNNI}(a) and (b) we report, as a function of $k_a$, the fraction
$S$ of nodes in the largest CMP cluster for $\gamma<3$ on uncorrelated
configuration model networks (UCM)~\cite{Catanzaro2005} of various size.  The
plot shows the presence of a CMPGC, including a fraction of active nodes
independent of the system size $N$.  The scaling of $S$ with $k_a$ is in
excellent agreement with the prediction of Eq.~(\ref{S}).  Finite size effects
are also apparent and perfectly agree with the prediction formulated above: The
effective threshold occurs for $k_a = \kmax(N) = N^{1/2}$.  Increasing the
network size, the effective threshold diverges: Asymptotically there is a giant
component for any $k_a>0$.

For $\gamma>3$ we have considered a hard cut-off  $M=N^{1/(\gamma-1)}$ for the
degree sequence generated in the UCM model, in order to avoid the possible
appearance of outliers having a degree much larger than the average $\kmax$
\cite{Boguna09}.
Panels (c) and (d) of Fig.~\ref{figNNI} show that also in this
case the fraction of active nodes in the CMPGC is extensive and its dependence
on $k_a$ is well described by Eq.~(\ref{S1}).  This confirms the depicted
scenario about the formation of an extensive CMPGC and points out that for the
sizes considered only the preasymptotic scaling regime $S_1$ is observed, while,
as expected, do not see any trace of the asymptotic behavior (for an infinite
network) $S = S_2 \approx k_a^{1-\gamma}$.

Concerning finite size effects, for $\gamma>3$ as only the first scaling regime
is observed, the condition setting the effective threshold is Eq.~(\ref{k_ac}).
A direct numerical verification of it for CMP is very hard, as practically all
non-isolated nodes are part of the CMPGC and finite clusters (upon which methods
to determine the position of the threshold are based) are extremely rare. An
indirect numerical verification is provided below in the application to the SIS
model.
The observation of the effective threshold associated to the second mechanism
[Eq.~(\ref{k_ac2})] is impossible in practice as it would require huge networks
of size larger than $N_2^*$.

The conclusion of our analysis is that, in different manners depending on
whether $\gamma<3$ or $\gamma>3$, a CMP giant component is present in
infinite networks for any value of $k_a$. The threshold for this class of CMP
processes is infinite for any value of $\gamma$.

\section{Application to SIS epidemic spreading}
\label{sec:appl-sis-epid}

The theoretical picture presented in the previous Sections can be applied to the
CMP process associated to SIS dynamics, which is an instance of this class with
$k_a=a/\lambda^2 \ln(1/\lambda)$, initial mass equal to the degree, and $r(m)=m/k_a$, see
Appendix~\ref{sec:conn-betw-cmp}.  This application has mainly the goal of
investigating the properties of the SIS epidemic transition for $\gamma>3$.
We notice that the relation between CMP and SIS depends on the
  parameter $a$ relating $k_a$ with $\lambda$, whose value, either $a=1$ or
  $a=4$, is not theoretically determined. In our application of CMP to SIS
we choose to compare with both values.

\subsection{Scaling of the CMP giant component}
\label{sec:scal-giant-comp}

The scaling of $S$ with $\lambda$ for $\gamma<3$
is obtained by inserting the expression for $k_a$ as a function of
$\lambda$ into Eq.~(\ref{S}), obtaining
\begin{equation}
  S = \frac{N_{a}}{N} \sim k_a^{1-\gamma} \sim \lambda^{2(\gamma-1)}
  \ln^{1-\gamma}\left(\frac{1}{\lambda}\right)
\end{equation}
Thus the approach predicts the existence of a CMPGC for any value of
$\lambda>0$.  Notice, however, that while it is possible to define a CMP process
associated to SIS dynamics for any value of $\gamma$, the SIS epidemic
transition for $\gamma<5/2$ is due to a mechanism different from the mutual
reinfection of distant hubs~\cite{Castellano2012}: Hence SIS critical properties
have nothing to do with those of CMP in this case.  Moreover, the connection
between the scaling of $S$ and the scaling of the SIS prevalence is not trivial
in this case, hence we cannot derive from CMP any prediction on the latter even
for $5/2 < \gamma <3$.

For $\gamma>3$ the fraction of active nodes in the CMPGC is extensive
and its preasymptotic dependence on $\lambda$ is obtained by plugging
the expression for $k_a$ into Eq.~(\ref{S1}):
\begin{equation}
  S_1 = \frac{N_{NI}}{N} \sim k_a^{2(2-\gamma)} \sim \lambda^{4(\gamma-2)}
  \ln^{2(2-\gamma)}\left(\frac{1}{\lambda}\right).
\end{equation}

We can also calculate the asymptotic scaling of the CMPGC, by plugging
the expression for $k_a$ into the expression of the scaling of the
CMP giant component in the second regime, Eq.~(\ref{S2}), obtaining
\begin{equation}
  S_2 \sim \ln^{1-\gamma}(k_a) k_a^{1-\gamma} \sim \lambda^{2(\gamma-1)}
  \ln^{2(1-\gamma)}\left(\frac{1}{\lambda}\right).
\end{equation}
We remind however, that this scaling occurs only for exceedingly large values of
$k_a$ (i.e., values of $\lambda$ exceedingly small), so that it cannot be
observed in present simulations.

\subsection{Finite-size epidemic threshold}
\label{sec:finite-size-epidemic}

\begin{figure}
  \includegraphics[width=\columnwidth]{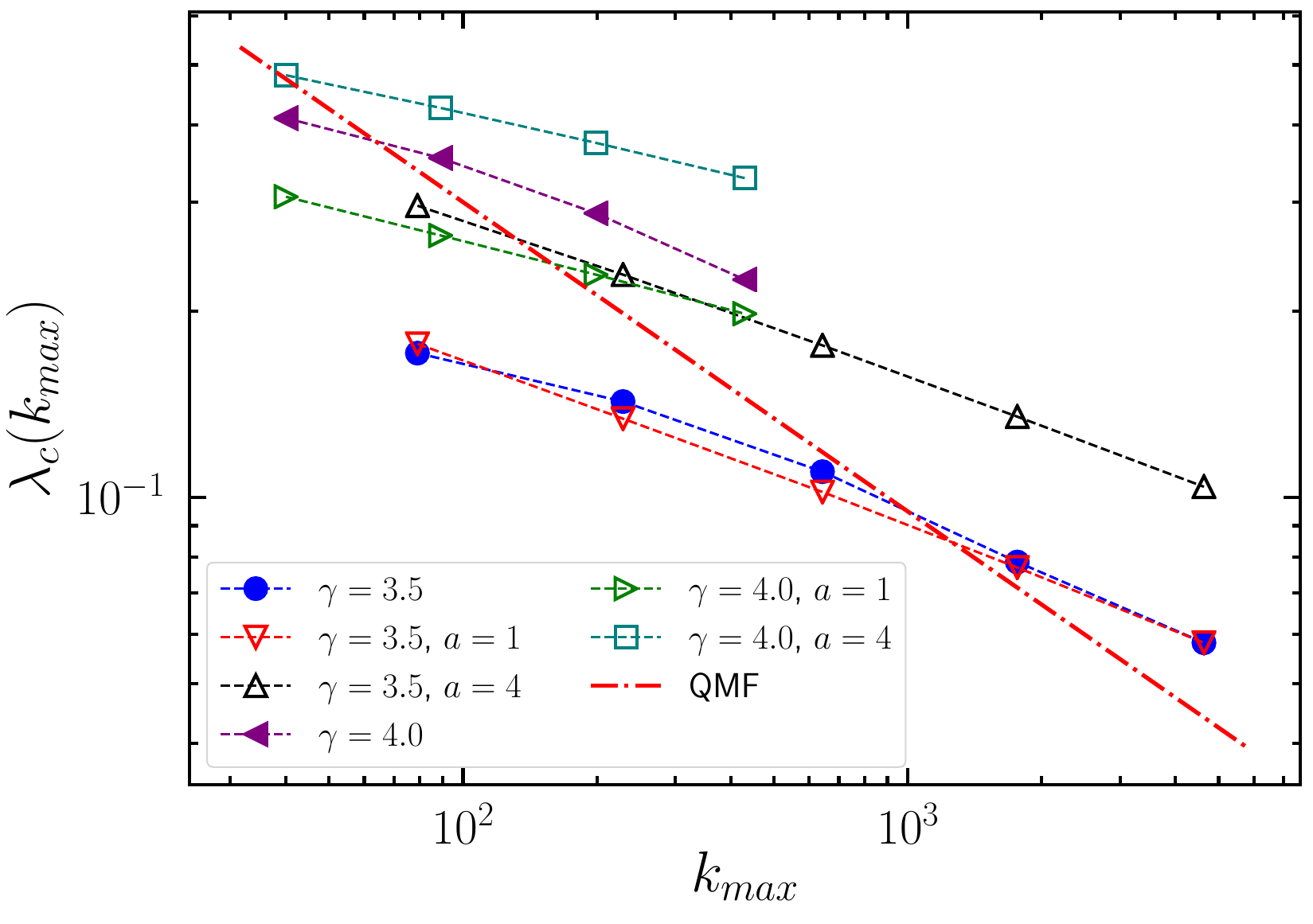}
  \caption{Comparison between the theoretical finite size threshold (for two
    different values of $a$), hollow symbols, and direct numerical simulations,
    full symbols, of the SIS process in UCM networks with degree exponent
    $\gamma=3.5$ ($\km=3$) and $\gamma = 4$ 
    ($\km=2$)~\cite{Boguna2013}. The epidemic threshold was
    determined by means of the lifespan method~\cite{Boguna2013}.
    The dashed red line is
    proportional to the prediction $1/\sqrt{\kmax}$ of QMF theory. 
    The values of $\kmax$ correspond to sizes ranging from 
      $N=10^4$ to $N=10^8$ for $\gamma=3.5$ and from $N=10^4$ to $N=10^7$ for
   $\gamma=4$.}
  \label{ThresholdComparison}
\end{figure}

For $\gamma>3$, as only the first scaling regime
is observed, the condition setting the effective threshold is
$\kmax(N)=k_c(\lambda)$, i.e.,
\be
\frac{a}{\lambda_c^2}\ln\left(\frac{1}{\lambda_c^2}\right) =
\km \kmax^{1/(\gamma-2)}.
\label{lambdaFS}
\ee
This translates (apart from logarithmic corrections)
into
\begin{equation}
  \lambda_{c}(N) = (a/\km)^{1/2} \kmax^{-1/[2(\gamma-2)]}.
\label{lambdaFS2}
\end{equation}
For $\kmax^{-1/2}<\lambda<\lambda_c(N)$ there are active hubs in the system, but
they do not give rise to a CMPGC.  Hence $\lambda_c(N)$ can be identified with
the effective size-dependent epidemic threshold.  Eq.~(\ref{lambdaFS2}) is very
interesting as it shows that the effective threshold does not vanish as
$\kmax^{-1/2}$, as predicted by QMF theory, but {\em more slowly}, with an
exponent that is reduced as $\gamma$ is increased.  The prediction of
Eq.~(\ref{lambdaFS}) is compared in Fig.~\ref{ThresholdComparison} with SIS
numerical results of Ref.~\cite{Boguna2013}, displaying a good agreement and
thus clarifying a long-standing open issue. For reference, 
we also report the scaling predicted by QMF theory, which patently 
disagrees with numerical results.

Notice, however, that this is not the final asymptotic behavior
of $\lambda_c(N)$. For much larger networks it could be possible
(at least in principle) to reach values of $k_a$ larger than the
crossover value $k_2^*$.
In such a case the decay of the effective threshold would be given
by the condition $k_x=\kmax(N)$, that, from Eq.~\eqref{eqomega}, 
leads to
\begin{equation}
  \lambda_c(N) = \omega^{1/2} \kmax^{-1/2}
  \sim \ln(\kmax) \kmax^{-1/2}.
\end{equation}
In this way we recover the asymptotic scaling of
the effective threshold recently derived by Huang and Durrett~\cite{Huang2018}.

In Appendix~\ref{SISstretched} we show that the CMP approach provides the
correct effective finite-size threshold also in the case of stretched
exponential degree distributions.

\subsection{SIS prevalence as a function of $\lambda$}
\label{sec:prev-as-funct}

Above the size-dependent effective threshold there is a backbone of active nodes
which sustain an endemic state by reinfecting each other.  In an infinite
network, when the CMP giant component is formed by distant, mutually interacting
hubs (second regime) we can estimate the value of the prevalence (average
density of infected nodes) for small $\lambda$ using the following argument.
All actives nodes with degree larger than $k_x=\omega k_a$ participate in the
CMPGC.  Each one of these active nodes of degree $k$ infects a number of other
nodes of order $\lambda k$. Since hubs are distant, these clusters of infected
nodes do not overlap, hence the total prevalence in the system is expected to
be~\cite{Lee2013}
\begin{equation}
  \label{eq:17}
  \rho \sim \int_{k_x}^{\infty} dk \,\lambda k \,P(k) \sim \lambda
  (\omega k_a)^{2-\gamma}.
\end{equation}
Substituting the values of $\omega$ and $k_a$ into Eq.~\eqref{eq:17} leads to
\begin{equation}
  \label{eq:20}
  \rho(\lambda) \sim \lambda^{2\gamma - 3} \left[ \ln(1/\lambda) \right]^{2(2-\gamma)},
\end{equation}
in agreement with the exact mathematical results of Mountford et
al.~\cite{Mountford2013}.  As discussed above, this prediction is, however,
impossible to verify numerically, because the onset of the asymptotic regime
could be seen only for exceedingly large networks.  This explains the mismatch
between the theory of Mountford et al. and numerical results.  In doable
simulations of the SIS model, the small $\lambda$ regime that can be observed is
the preasymptotic regime $S_1$ for the corresponding CMP process. In such a
regime, since hubs are not well separated, it is not possible to assume that
each of them independently infects a number of neighbors of the order of
$\lambda k$.  The derivation of the exponent characterizing the SIS prevalence
singularity in this preasymptotic (but long) regime remains an interesting open
question for future research.

\section{Discussion}
\label{sec:discussion}

In this paper we have considered a long-range percolation process, the
cumulative merging percolation, exhibiting a rich phenomenology that we have
uncovered developing an appropriate scaling theory. While we have mainly focused
on particular forms of the model inspired by the analysis of SIS
process~\cite{Menard2015}, more complex scenarios can be obtained by changing
the functional form of the interaction range $r(m)$, the activation probability
$p_i$ and by considering a more complicated mass merging function
$m_{\alpha + \beta}=g(m_\alpha,m_\beta)$. In this sense, we expect other types
of percolation transitions to arise as these features are changed. For example,
if $r(m)$ saturates to a finite value when $m$ diverges, the arguments presented
above imply the presence of a finite threshold for $\gamma>3$ as for the DOP
process.  The investigation of the general phenomenology of the CMP process and
of its connections with other models is a promising avenue for future research.

Concerning the application of CMP to SIS dynamics, our results 
clarify how the mutual reinfection mechanism among distant hubs, underlying
the epidemic transition for $\gamma>3$~\cite{Boguna2013}, takes place. 
This closes the last
gap in our understanding of the SIS dynamics and leads to a complete and
consistent physical picture, that we sketch here.

\begin{figure}
  \includegraphics*[width=\columnwidth]{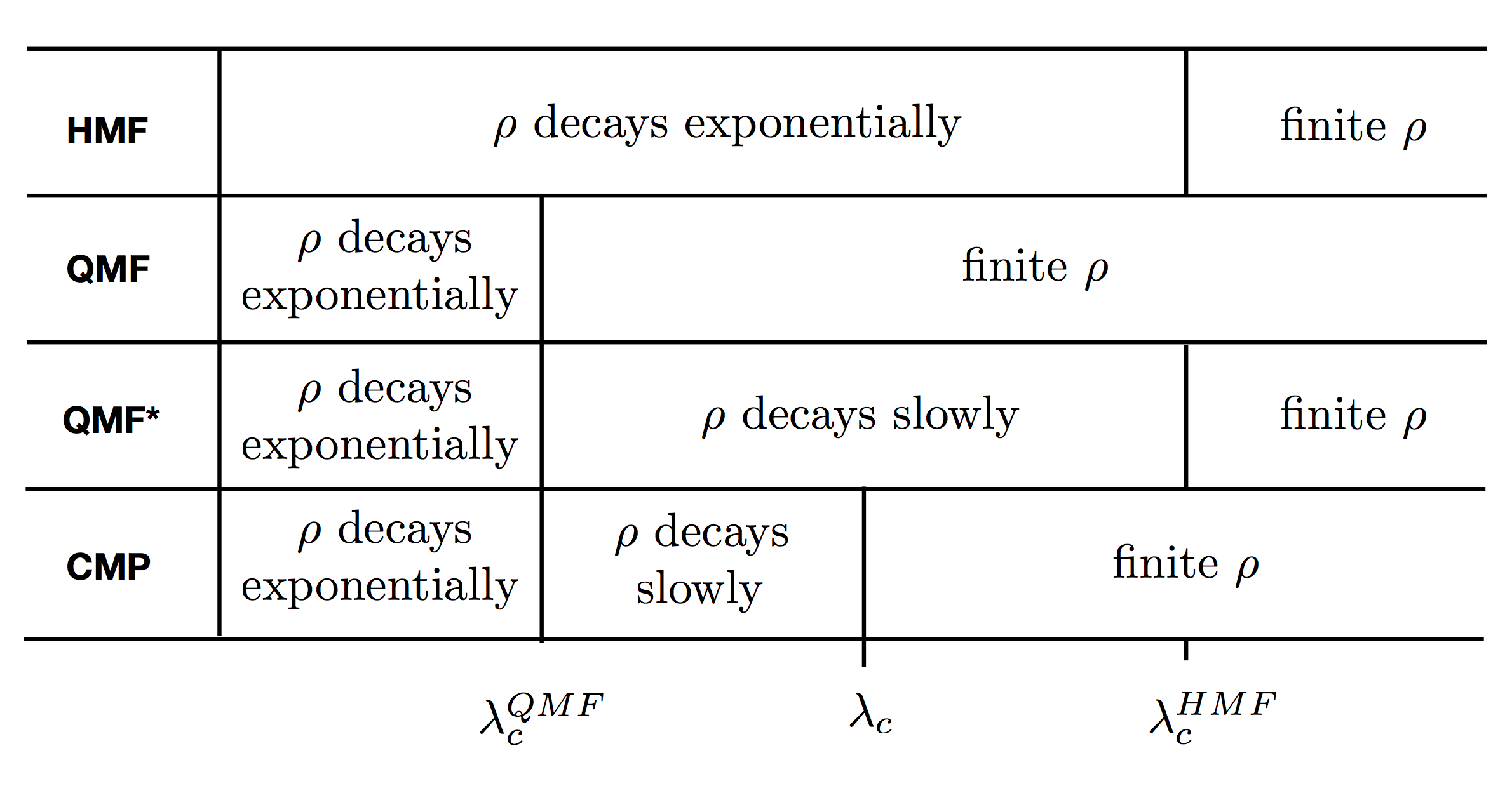}
  \caption{Behavior of SIS prevalence $\rho$ according to the different
    approaches. QMF* stands for the QMF theory as reinterpreted in 
Refs.~\cite{Goltsev2012,Lee2013}}.
\label{phasediagram}
\end{figure}

The original Heterogeneous-Mean-Field theory (HMF)~\cite{pv01a,Pastor01b}, 
based on an annealed network approximation~\cite{dorogovtsev07:_critic_phenom},
predicts an epidemic threshold given $\lambda_c^{HMF} = \av{k}/\av{k^2}$, and
thus finite in the limit of infinite size networks for $\gamma>3$ (see
Fig.~\ref{phasediagram}). Below  $\lambda_c^{HMF}$, this theory 
predicts a density of infected individuals $\rho(t)$
decaying exponentially to zero. 
For  $\lambda > \lambda_c^{HMF}$, HMF predicts a finite $\rho$ in the 
steady state.

Quenched-Mean-Field theory (QMF or
NIMFA)~\cite{Wang03,PVM_ToN_VirusSpread,Gomez10} predicts the same scenario,
a transition separating an active steady state with finite prevalence
from an absorbing phase where the prevalence decays exponentially to zero.
The difference with respect to HMF is in the value of the threshold,
$\lambda_c^{QMF}=1/\Lambda_M$, where $\Lambda_M$ is the largest eigenvalue of
the adjacency matrix, which vanishes as $N$ diverges, for any $\gamma$.
The value of the QMF threshold is
the minimum value of $\lambda$ such that the star graph composed
by the largest hub and its direct neighbors is able to 
independently sustain long-lasting 
activity~\cite{Castellano2012}. 
Correspondingly, the principal 
eigenvector for $\gamma>5/2$ is localized around the largest 
hub~\cite{Goltsev2012}.
Also the other leading eigenvalues of the adjacency matrix
are associated to eigenvectors localized on each of the hubs
of the network.
In the thermodynamic limit, for any given value of $\lambda$,
each large hub with degree $k > 1/\lambda^2$ sustains long-lasting
activity together with its direct neighbors, yielding an
overall finite density of infected individuals that can be
estimated~\cite{Lee2013} as $\rho \sim \lambda^{2\gamma - 3}$.

However, as pointed out in Refs.~\cite{Goltsev2012} and~\cite{Lee2013},
this scenario cannot really hold for SIS dynamics.
In QMF theory there are no stochastic fluctuations and activity
in a star graph composed by $k+1$ nodes persists forever if 
$\lambda>1/\sqrt{k}$.
In SIS dynamics instead, activity survives, in a star graph made of $k+1$ 
nodes, only for a time of the order of $\exp(\lambda^2 k/a)$.
Hence if star graphs are independent the overall activity
does not survive for a time scaling exponentially with the system 
size $N$. In other words, there is some activity surviving for some time
but not a truly steady active state.
In the interval $\lambda_c^{HMF} < \lambda < \lambda_c^{HMF}$ 
one should expect a Griffiths-like phase, with a 
slow decay of $\rho(t)$~\cite{Lee2013}, due to the convolution of 
contributions from a decreasing number of still active individual hubs.
Only above a {\em finite} threshold, approximately equal to $\lambda_c^{HMF}$, 
corresponding to the inverse of the first eigenvalue associated 
to a delocalized eigenvector~\cite{Goltsev2012}, 
a truly active steady state is expected (see Fig.~\ref{phasediagram}).
This scenario is consistent, but at odds with numerical 
simulations~\cite{Boguna2013} 
and exact mathematical results~\cite{Chatterjee2009,Mountford2016},
which find an active steady state for any value of $\lambda>0$ in the
thermodynamic limit.

One possible way to reconcile these findings was explored by Lee et 
al.~\cite{Lee2013}. If large hubs were in mutual direct contact
(i.e., they form an extensive connected cluster) should
activity spontaneously disappear in one of them neighboring hubs 
would be able to reinfect it; these mutual reinfections would lead
to a survival time exponential in $N$, i.e., a truly active steady state.
Unfortunately, the study of Degree-Ordered-Percolation revealed
that such an extensive cluster of large hubs always exists only for $\gamma<3$.
For $\gamma>3$ the DOP threshold is finite: the largest hubs
are separated and do not form an extensive cluster~\cite{Lee2013}.

Our results go beyond those of Lee et al. and clarify what was missing 
in previous approaches: the activity triggered by hubs extends beyond 
nearest neighbors, up to a scale  that grows with $\lambda$, 
so that hubs can interact even if they are not in direct contact.
This long-range interaction gives rise, above a critical value $\lambda_c(N)$, 
to an extensive CMP percolating cluster of active nodes able to 
reinfect each other at distance and thus giving a veritable steady state
with finite prevalence $\rho$.  
The threshold $\lambda_c(N)$ is intermediate between $\lambda_c^{QMF}$
and $\lambda_c^{HMF}$ and vanishes as a function of $N$ (at odds with
$\lambda_c^{HMF}$) but more slowly than $\lambda_c^{QMF}$.
Considering finite networks, while for 
$\lambda_c(N) < \lambda < \lambda_c^{HMF}$ a CMP percolating cluster
exists and prevalence is finite, for $\lambda_c^{QMF} < \lambda < \lambda_c$ 
only small nonpercolating CMP clusters are present. 
In this case each of them decays independently and thus a Griffiths-like phase,
characterized by $\rho(t)$ slowly decaying to zero, is expected
(Fig.~\ref{phasediagram}). 
A numerical validation of
this prediction, which is difficult as both interval bounds vanish with the
system size, remains a challenge for future numerical studies.

The consideration of long-range effects is the crucial ingredient
in our analysis that makes a qualitative difference with previous
approaches. While QMF theory neglects correlations among the dynamical 
state of neighbors, other theories~\cite{Mata2013, Gleeson11, StOnge2018} 
take some correlations into account, but since they consider only 
neighbors in a short range, they could not capture the long-range
percolative nature of the SIS epidemic transition for $\gamma>3$.

Our work puts in proper place the different theories presented in recent years
to explain the behavior of the SIS model in heterogeneous networks, showing in
particular the limit in which exact mathematical results are expected to be
observed, putting thus an end to the long debate on this subject.
On the other hand it opens new perspectives, as it proposes the
cumulative merging of distant clusters as a very generic
phenomenon which may originate nontrivial types of percolation phenomena
in networks.

\begin{acknowledgments}
  We acknowledge financial support from the Spanish Government's MINECO, under
  project FIS2016-76830-C2-1-P.  R. P.-S. acknowledges additional financial
  support from ICREA Academia, funded by the \textit{Generalitat de Catalunya}
  regional authorities.
\end{acknowledgments}

\appendix

\section{Connection between CMP and SIS}
\label{sec:conn-betw-cmp}

The Susceptible-Infected-Susceptible (SIS) model, often called contact process
in the community of applied probabilists, is defined as follows: Individuals can
be in one of two states, either susceptible or infected. Susceptible individuals
become infected by contact with infected individuals, at a rate equal to the
number of infected contacts times a given spreading rate $\beta$. Infected
individuals on the other hand become spontaneously healthy again at a rate
$\mu$.  The ratio $\lambda=\beta/\mu$ is the control parameter for the model,
which experiences a transition between a healthy and an endemic (infected)
steady state when $\lambda$ crosses an epidemic threshold $\lambda_c$. In
power-law distributed networks, for $\gamma>5/2$ the epidemic transition is
triggered by nodes with a large number $k$ of neighbors (hubs).  Each of these
hubs together with its direct neighbors (leaves) forms a star graph, which in
isolation is able to sustain the survival of the epidemic for a long time,
$\tau(k) \sim \exp(\lambda^2 k/4)$~\cite{Boguna2013}, provided $\lambda$ is
larger than $\lambda_c(k)=1/\sqrt{k}$.  During this long time interval, even if
the hub recovers from the infection, it is promptly reinfected by one of its
neighbors, and can in its turn reinfect other leaves when they recover.  After a
typical time $\tau$ a fluctuation takes the star graph formed by a hub and its
nearest neighbors to the absorbing state.

Since the star graph is not isolated in the network, it can propagate activity
to other nodes. It is possible to estimate~\cite{Boguna2013} the average
time it takes for an infected node to infect for the first time a node at
distance $r$ in the limit of small $\lambda$,
\begin{equation}
  \label{eq:1}
  T(r) \sim e^{r\ln(1/\lambda)}.
\end{equation}
By equating $\tau$ and $T(r)$ it is possible to estimate the ``range of
interaction'' of a hub of degree $k$, i.e., the maximum distance at which a star
surrounding an active hub is able to propagate the infection before
spontaneously recovering:
\begin{equation}
  r(k) = \frac{\lambda^2k}{4 \ln(1/\lambda)} = \frac{k}{k_a},
\end{equation}
where we have defined
\begin{equation}
  k_a = 4 \frac{1}{\lambda^2} \ln\left(\frac{1}{\lambda}\right).
\end{equation}

\begin{figure}[t]
  \includegraphics*[width=\columnwidth]{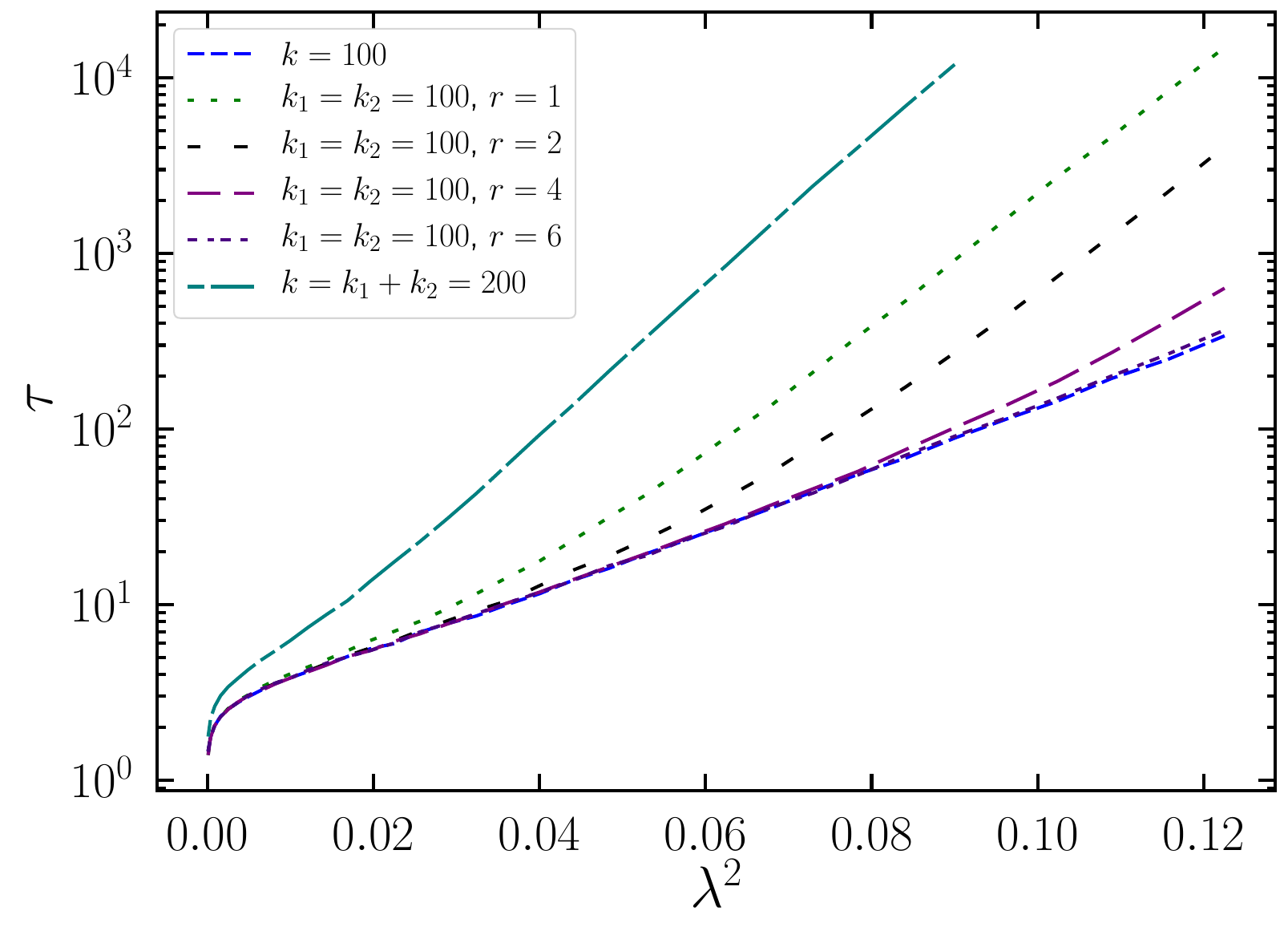}
  \caption{Average survival time $\tau$ of a SIS epidemic on two star graphs of
    size $k_1 = k_2 = k$, connected by a line of $r-1$ intermediate nodes,
    starting with only the hub of one of them in the infected state. We compare
    with the result for single stars of size $k$ and $2k$. As we can see, for
    sufficiently small values of $r$ (distance between the hubs), the survival
    time of the connected stars of size $k$ scales with $\lambda^2$ as a single
    star of size $2k$.}
  \label{figS2}
\end{figure}

Consider now another hub, of degree $k'$ at a distance $r_0$ from the first.  If
$r_0<r(k)$ the second hub will be infected by the first and it will be able to
stay infected (together with its direct neighbors) for a time $\tau(k')$.
During this time interval it will spread the infection up to a distance
$r(k')$. If $r_0>r(k')$ this means that the second hub will not be able to
reinfect the first, should it have fallen into the absorbing state.
Conversely, if $r_0<r(k')$, even if the first hub recovers, it will be
reinfected by the second. In this way, the two distant hubs form a coupled
system such that if one hub recovers the other is able to reinfect it before
recovering in its turn. For the infection to die out in the system of the
two hubs, they must recover almost simultaneously~\cite{Menard2015}.
This happens after a time of the
order of $\tau(k) \tau(k') \sim \exp[\lambda^2 (k+k')]$, see Fig.~\ref{figS2}.

Hence the combined set of hubs will be able to infect nodes at an increased
range of interaction $r(k+k')$. It is then clear that SIS dynamics can be seen
as an instance of the Cumulative Merging Process, with active nodes those with
$k \ge k_a=4/\lambda^2 \ln(1/\lambda)$, initial masses equal to node degrees
$m_i^{(0)}=k_i$ and range of interaction given by $r(m) = m/k_a$.
Notice that the factor 4 in the expression for $k_a$ is the
consequence of the choice $\tau(k) \sim \exp(\lambda^2 k/4)$.  Alternative
treatments~\cite{Cator2013,Huang2018} give that star graphs are active for
$k> k_a = 1/\lambda^2 \ln(1/\lambda)$.  In the comparison of the CMP approach to
SIS with numerical simulations we consider both expressions.

\section{Average distance between a node of degree $k$ and the closest node of
  degree at least $k$}
\label{sec:aver-dist-betw}

We can estimate the average distance $d(k)$ between a node of degree $k$ and the
nearest node of degree larger than or equal to $k$ within a
tree-like approximation~\cite{dorogovtsev07:_critic_phenom}. For random
uncorrelated networks, the probability that a link points to a node of degree
$k'$ is $k' P(k') / \av{k}$. Arriving at this node, there are $k' - 1$ possible
outgoing edges (excluding the one used to arrive to node $k'$). The average
number of outgoing edges (the so-called branching factor) is thus
\begin{equation}
  \label{eq:8}
  \kappa = \int_{\km}^\infty (k'-1)\frac{k' P(k')}{\av{k}} dk' = \frac{\av{k^2}}{\av{k}} -
  1,
\end{equation}
that is a finite number for power-law networks with $\gamma > 3$. From this
branching ratio, we estimate the average number of nodes at distance $n$ as
$N_n = k \kappa^{n-1}$, assuming the tree approximation.

A node of degree $k$ has $k$ neighbors. It is connected at distance $d=1$ to a
node of degree not less than $k$ if at least one of these neighbors has degree
larger than or equal to $k$. The probability of this event is
\begin{equation}
  \label{eq:9}
  P_>(k) = \int_{k}^\infty \frac{k' P(k')}{\av{k}} dk'= \left( \frac{k}{\km}
  \right)^{2-\gamma}.
\end{equation}
Therefore, the probability that the distance at the nearest neighbors with
degree larger than or equal to $k$ is equal to $d=1$ is
\begin{equation}
  \label{eq:10}
  P(d=1) =  1 - \left[ 1 -  P_>(k) \right]^k = 1 - \left[  P_<(k) \right]^k,
\end{equation}
where $P_<(k) = 1 - P_>(k)$ is the probability that a nearest neighbor of a node
has degree smaller than $k$.

The nearest neighbor with degree not less than $k$ is at distance $d=2$ if there
are no such neighbors at distance $d=1$, and at least one of the neighbors at
distance $d=2$, in number $k \kappa$, has degree not less than $k$, which
happens with probability
\begin{equation}
  \label{eq:11}
  P(d=2) = \left[  P_<(k) \right]^k \left[  1 - \left[ P_<(k) \right]^{k
      \kappa} \right].
\end{equation}
By induction, we can see that the nearest neighbor of degree not less that $k$
is at distance $d=n$ corresponds to not observing one at any distance smaller
than $n$, and having at least one at distance equal to $n$, which happens with
probability
\begin{widetext}
\begin{eqnarray}
  \label{eq:12}
  P(d=n) &=&  \left[ P_<(k) \right]^k  \left[  P_<(k) \right]^{k \kappa}
  \left[ P_<(k) \right]^{k \kappa^2} \cdots  \left[  P_<(k) \right]^{k
    \kappa^{n-2}} \left[  1 - \left[ P_<(k) \right]^{k
  \kappa^{n-1}} \right] \nonumber \\
  &=& \left[  P_<(k) \right]^{\sum_{r=0}^{n-2} k \kappa^r} - \left[   P_<(k)
      \right]^{\sum_{r=0}^{n-1} k \kappa^r} \nonumber\\
         &=& \left[ P_<(k) \right]^{k(\kappa^{n-1}-1)/(\kappa -1)} -
             \left[ P_<(k) \right]^{k(\kappa^{n}-1)/(\kappa -1)} \nonumber \\
  &=& \frac{\left[ P_<(k) \right]^{k \kappa^{n-1}/(\kappa -1)} -
             \left[ P_<(k) \right]^{k\kappa^{n}/(\kappa -1)}}{ \left[ P_<(k)
      \right]^{k/(\kappa -1)}} \equiv \frac{C^{\kappa^{n-1}} - C^{\kappa^n}}{C},
\end{eqnarray}
\end{widetext}
where for simplicity we set $C = \left[ P_<(k) \right]^{k/(\kappa -1)}$.

The average distance $d(k)$ can be evaluated as
\begin{eqnarray}
  d(k)& =& \sum_{n=1}^\infty n P(d=n) = \sum_{n=1}^\infty n \frac{C^{\kappa^{n-1}} -
          C^{\kappa^n}}{C} \nonumber \\
  &=& \sum_{n=0}^\infty \frac{C^{\kappa^n}}{C} \label{eq:13}.
\end{eqnarray}

The summation in Eq~\eqref{eq:13} cannot be performed directly.
We can approximate its
behavior for large $k$ by transforming it into an integral:
\begin{eqnarray}
   d(k) &\simeq& \frac{1}{C} \int_0^\infty C^{\kappa^x} dx = \frac{1}{C \ln(\kappa)}
                 \int_1^\infty \frac{C^z}{z} dz \\ \nonumber
  &=&  \frac{\Gamma(0, -\ln(C))}{C \ln(\kappa)},\label{eq:14}
\end{eqnarray}
where $\Gamma(a, z)$ is the incomplete Gamma function~\cite{abramovitz}, and we
have applied the change of variables $\kappa^x = z$. For large $k$,
$C = \left[ 1 - \left( \frac{k}{\km}\right)^{2-\gamma} \right]^{k/(\kappa
  -1)}$ tends to $1$, so we can expand the incomplete Gamma function in
Eq.~\eqref{eq:14} for small arguments,
$\Gamma(0,z) \sim -\ln(z)$~\cite{abramovitz}, to obtain the asymptotic behavior
\begin{equation}
  \label{eq:15}
  d(k) \sim \frac{-\ln[-\ln(C)]}{C \ln(\kappa)} \sim \frac{\ln \left[
      \left(  \frac{k}{\km} \right)^{\gamma-3} \frac{\kappa -1}{\km}
    \right]}{\ln(\kappa)},
\end{equation}
where we have expanded $C$ for large $k$. We therefore observe the asymptotic
behavior for large $k$ in infinite networks as
\begin{equation}
  \label{eq:16}
  d(k) \sim 1 + \frac{\gamma-3}{\ln(\kappa)} \ln \left( \frac{k}{\km} \right),
\end{equation}
where the term $1$ accounts for the minimum possible distance between nodes.

\begin{figure}[t]
  \includegraphics*[width=\columnwidth]{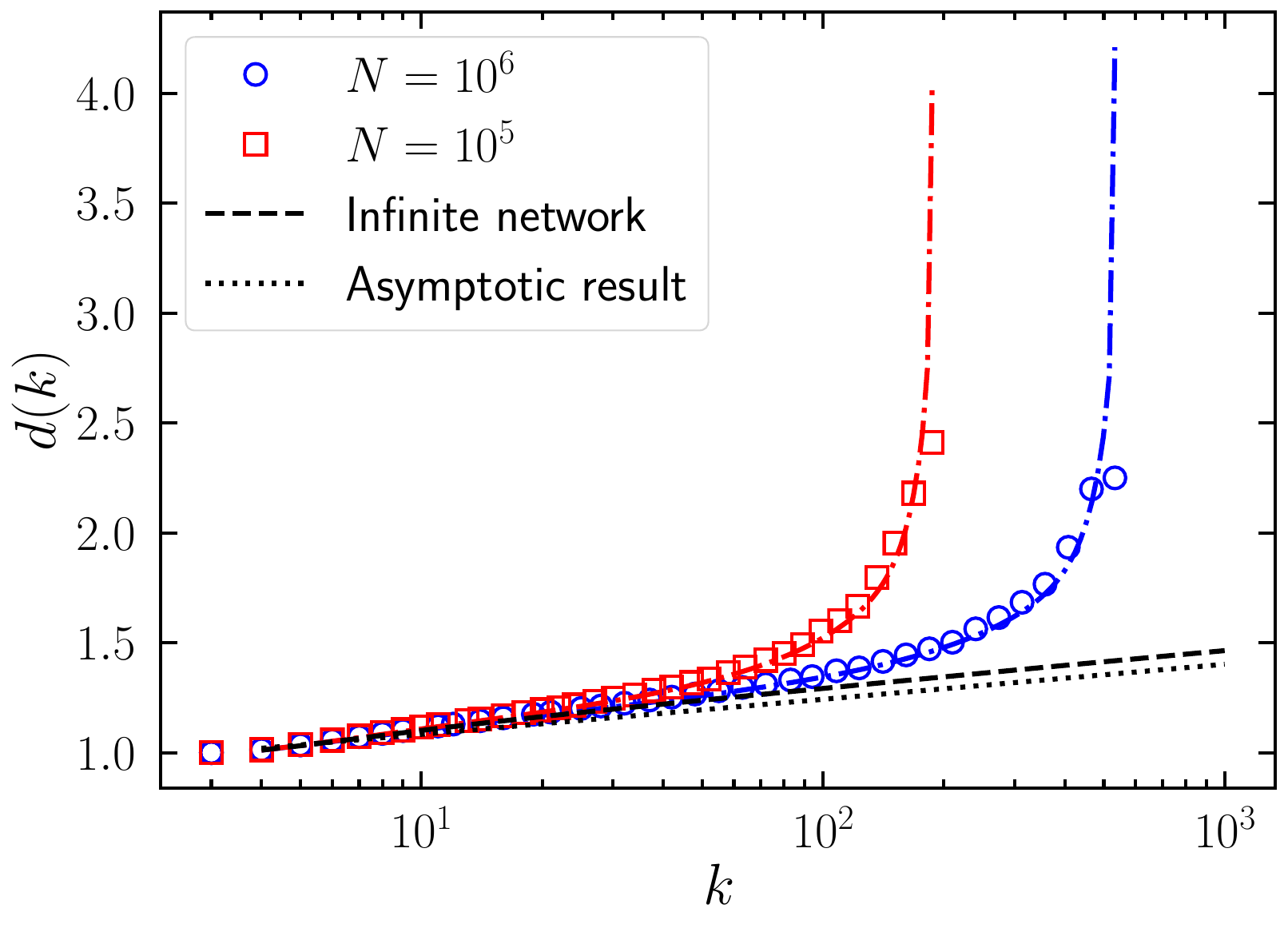}
  \caption{Average distance between a node of degree $k$ and the closest node
    of
    degree at least $k$ in power-law networks with degree exponent
    $\gamma=3.2$. Symbols represent numerical simulations over networks of
    different size, averaged over $10^4$ independent network samples. The dashed
    line represents the numerical evaluation of the summation Eq.~\eqref{eq:13}
    in the infinite network limit. Dot-dashed lines represent results of the
    summation for finite networks of the corresponding size $N$. The dotted line
    corresponds to the asymptotic expression Eq.~\eqref{eq:16}.}
  \label{figSSS}
\end{figure}

This calculation, performed in the tree approximation, captures nevertheless the
behavior in real uncorrelated power-law networks generated with the Uncorrelated
Configuration Model (UCM)~\cite{Catanzaro2005}. In Fig.~\ref{figSSS} we present
the result of numerical simulations, together with the numerical evaluation of
the summation in Eq.~\eqref{eq:13}, performed using a discrete power-law degree
distribution $P(k) = k^{-\gamma}/[\zeta(\gamma, \km) - \zeta(\gamma, \kmax)]$,
where $\zeta(s,a)$ is the Hurwitz zeta function~\cite{abramovitz}. The dashed
line represents the result for an infinite network ($\kmax = \infty$), while the
dot-dashed lines mark the value for networks with maximum degree
$\kmax = N^{1/(\gamma-1)}$~\cite{mariancutofss}. The dotted line shows the
asymptotic behavior obtained in Eq.~\eqref{eq:16}.

\section{Cumulative Merging Percolation on stretched exponential networks}
\label{stretched}

Let us consider the example of a network with cumulative degree
distribution~\cite{Huang2018}
\begin{equation}
  \label{eq:1s}
  P_c(k) = e^{-k^{\beta} + \km^{\beta}},
\end{equation}
corresponding to a stretched exponential degree distribution
\begin{equation}
  \label{eq:2s}
  P(k) = - \frac{d P_k(k)}{d k} = \beta  k^{\beta -1} e^{\km^{\beta }-k^{\beta }}.
\end{equation}
Applying extreme value theory, for a finite network of size $N$ we have
\begin{equation}
  \label{eq:8s}
  \kmax \sim [ \ln(N)]^{1/\beta}.
\end{equation}

The other relevant quantities for CMP are
\begin{equation}
  \label{eq:4s}
  \frac{N_a}{N} =   \int_{k_a}^{\infty}  dk  \;P(k) = e^{\km^{\beta }-k_a^{\beta }}
\end{equation}
and
\begin{eqnarray}
  \label{eq:3}
  P_a =  \frac{1}{k_c} &=& \int_{k_a}^{\infty}  dk  \;\frac{k P(k)}{\av{k}} =
  \frac{\Gamma \left(1+\frac{1}{\beta },k_a^{\beta }\right)}{\Gamma
          \left(1+\frac{1}{\beta },\km^{\beta }\right)} \\
  &\simeq& \frac{e^{-k_a^\beta} k_a}{\Gamma
          \left(1+\frac{1}{\beta },\km^{\beta }\right)},
\end{eqnarray}
where we have developed the numerator in the limit of large $k_a$.
The average number of active neighbors for each active node is
\begin{eqnarray}
  \label{eq:5s}
  \frac{N}{N_a} \int_{k_a}^{\infty}  dk \; k P(k) P_a &=&
\frac{e^{k_a^{\beta}} \Gamma \left(1+\frac{1}{\beta },k_a^{\beta
    }\right)^2}{\Gamma \left(1+\frac{1}{\beta
    },\km^{\beta }\right)} \nonumber \\
  &\sim&  \frac{e^{-k_a^\beta}k_a^{2}}{\Gamma \left(1+\frac{1}{\beta },\km^{\beta
    }\right)},
\end{eqnarray}
where we have expanded the last expression in the limit of large
$k_a$. Therefore the average number of active neighbors
of an active node vanishes exponentially.
This implies that the size of the DOPGC decays exponentially fast
and the extended DOP mechanism is not at work: Small clusters
are at distance much larger than 2 from the DOPGC.

The only mechanism leading to the formation of the CMPGC is the second
one, based on the interaction at distance among isolated nodes.
This involves a scale $k_x = \omega k_a$, such that
$r(k_x) = d(k_x)$, to ensure that all nodes with $k>k_x$ see each other
and can merge in the same cluster.
To compute $d(k)$, from Appendix~\ref{sec:aver-dist-betw} we must
evaluate, in the limit of large $k$,
\begin{equation}
  \label{eq:9s}
  d(k) \sim \frac{-\ln\left[ -\ln(C) \right]}{C \ln(\kappa)},
\end{equation}
with $C = \left[  P_<(k) \right]^{k/(\kappa -1)}$ and $\kappa$ the branching
factor. In this case
\begin{equation}
  \label{eq:10s}
  P_<(k) = 1 - P_c(k) = 1 - e^{\km^{\beta }-k^{\beta }}.
\end{equation}
For large $k$,
\begin{equation}
  \label{eq:11s}
  -\ln(C) \simeq \frac{k}{\kappa-1}e^{\km^{\beta }-k^{\beta }},
\end{equation}
and
\begin{equation}
  \label{eq:12s}
  -\ln\left[ -\ln(C) \right] \simeq k^{\beta } - \km^{\beta } - \ln \left(
    \frac{k}{\kappa-1}  \right).
\end{equation}
Therefore, for large $k$,
\begin{equation}
  \label{eq:13s}
  d(k) \simeq \frac{k^\beta}{\ln(\kappa)},
\end{equation}
where we have disregarded constant and logarithmic terms.
For $r(k) = k/k_a$, from $d(k_x) = r(k_x)$, we obtain
\begin{equation}
  \label{eq:15s}
  \omega =  \frac{ \omega^\beta k_a^\beta}{\ln(\kappa)},
\end{equation}
leading to
\begin{equation}
  \label{eq:16s}
  \omega = \left( \frac{k_a^{\beta} }{\ln(\kappa)}
  \right)^{1/(1-\beta)}.
\end{equation}
So, we have
\begin{equation}
  \label{eq:17s}
  k_x = k_a \omega =  \left( \frac{ k_a }{\ln(\kappa)}
  \right)^{1/(1-\beta)}.
\end{equation}
As a consequence
\begin{eqnarray}
S_2 &=&  \int_{k_x}^{\infty}  dk  \;P(k) = e^{\km^{\beta }-k_x^{\beta }} \\
& \approx & e^{\km^{\beta }-(k_a/\ln(\kappa))^{\beta/(1-\beta)}}.
\end{eqnarray}

\section{Application to SIS on stretched exponential networks}
\label{SISstretched}

Since the asymptotic behavior of the order parameter for
the CMP transition is given by $S_2$
the effective finite-size threshold is given by the condition
$k_x \simeq \kmax$, that is,
\begin{equation}
  \label{eq:18s}
  k_a  \simeq \ln(\kappa)  \kmax^{1-\beta}.
\end{equation}
Using $k_a = a (1/\lambda)^2 \ln(1/\lambda)$ this implies
\begin{equation}
  \label{eq:19s}
  a (1/\lambda_c)^2 \ln(1/\lambda_c)  \simeq \ln(\kappa) \kmax^{1-\beta}.
\end{equation}
Disregarding logarithmic factors, this expression can be inverted, leading, in
the limit of large $\kmax$, to
\begin{equation}
  \label{eq:20s}
  \lambda_c \simeq \sqrt{\frac{a(1-\beta)}{2\ln(\kappa)}} \left[
    \kmax^{\beta-1} \ln(\kmax) \right]^{1/2}.
\end{equation}
For a stretched exponential degree distribution,
$\kmax \simeq \left[ \ln(N) \right]^{1/\beta}$, so we finally have
\begin{equation}
  \label{eq:21s}
  \lambda_c \simeq \sqrt{\frac{a(1-\beta)}{2\beta \ln(\kappa)}}
  \left[ \ln(N) \right]^{(\beta -1)/(2\beta)} \left[ \ln(\ln(N)) \right]^{1/2}.
\end{equation}
In this way we recover the exact logarithmic dependence of the
effective threshold on $N$ for the stretched exponential case,
recently found in Ref.~\cite{Huang2018}.

In the limit of a pure exponential distribution, $\beta=1$, the previous
arguments cannot be applied. However, recent results in
Ref.~\cite{Bhamidi2019} show that the threshold in this case is finite.

\input{CMP_resubmission2_upload.bbl}

\end{document}

%% file: CMP_resubmission2_upload.bbl
%